\documentclass[envcountreset]{llncs}
\usepackage{amssymb}
\usepackage{mathtools}
\begin{document}
\title{Bilinear cryptography using groups of nilpotency class 2\thanks{This research was supported by a NBHM and a SERB research grant.}}
\titlerunning{Bilinear cryptography and nilpotent groups}
\author{Ayan Mahalanobis \and Pralhad Shinde}
\authorrunning{Mahalanobis\and Shinde}
\institute{Indian Institute of Science Education and Research Pune\\Dr.~Homi Bhabha Road, Pashan, Pune, INDIA\\
\email{ayan.mahalanobis@gmail.com}}
\maketitle
\begin{abstract}
In this paper, we develop a novel idea of a bilinear cryptosystem using the discrete logarithm problem in matrices. These matrices come from a linear representation of a group of nilpotency class 2. We discuss an example at the end.
\end{abstract}
\section{Introduction}
Simply stated, this paper is an application of nilpotency class 2 groups in bilinear public-key cryptography to build a secure bilinear cryptosystem. Bilinear or pairing based cryptosystems are used in many practical situations such as the following:
\begin{itemize}
\item Identity based encryption: In this case the user's public-key is based
on his own identity, like his email address or phone number, see~\cite{palash}.
\item Short signatures: Signature schemes where the signature is short,
about half the size of the original signatures, see~\cite{boneh}.
\item Key exchange: Tripartite Diffie-Hellman key exchange~\cite{joux,indo}.
\item And others.
\end{itemize}
We are not going to survey all of pairing-based cryptographic protocols but
will refer the reader to~\cite{ratna}. However, we briefly talk about the tripartite Diffie-Hellman key exchange protocol purely as a motivation to our paper.

\section{A brief introduction to bilinear public-key cryptography}
The origin of pairing based cryptosystems is in the MOV attack~\cite{MOV} on
the elliptic curve discrete logarithm problem. The attack was first envisioned by Gerhard Frey. 
The idea was to use the bilinear properties of the
Weil pairing to reduce a discrete logarithm problem in an elliptic curve over
a finite field $\mathbb{F}_q$ to a discrete logarithm problem in $\mathbb{F}_{q^k}$. It is known~\cite{balu} that
most of the time for non super-singular curves, this $k$, the embedding degree
is very large.

The Weil or the Tate pairing is a bilinear map
\[B\;:G\times G\rightarrow\mathbb{F}_{q^k}.\]
Where $G$ is the (abelian) group of the elliptic curve written additively. Then
the discrete logarithm problem in $G$ to the base $g\in G$ is given $g$ and $\alpha g$ find
$\alpha$. It was Joux~\cite{joux} who first noticed that one can manipulate the bilinear
map to deliver a one-round tripartite Diffie-Hellman key exchange protocol.
The idea of using a bilinear map can also be traced back to the work of Dan
Boneh on the decisional Diffie-Hellman problem.
Let $\mathcal{A}, \mathcal{B}$ and $\mathcal{C}$ be three users who want to set up a common secret key
among themselves. Then choose three integers $\alpha$, $\beta$ and $\gamma$ respectively and
keep it a secret. They then compute $\alpha g$, $\beta g$ and $\gamma g$ respectively from the
public information $G = \langle g\rangle$ and broadcast this information over the public
channel. The user $\mathcal{A}$ on receiving $\beta g$, $\gamma g$ can compute $B(\beta g, \gamma g)^\alpha$ using his
private key $\alpha$. The same thing can be computed by $\mathcal{B}$ and $\mathcal{C}$ by using the
public information of the other two users and his private information. The
common key becomes $B(g, g)^{\alpha\beta\gamma}$. All is well and nice in what we just said,
except that $B$ being a alternate(skew-symmetric) map, $B(g, g) = 1$. There
are many approaches to solve that problem, one was proposed by Joux~\cite{joux}
and the other using a distortion map. In the interest of brevity of this paper we won't go into further details of pairing based cryptosystems using
elliptic curve. We will just have to comment on a few things. There are lots of issues with elliptic curve pairing. The most important of those are, how
to find curves with right embedding degree and what is the right embedding degree?

\section{A brief introduction to nilpotency class 2 groups and commutator identities}
For any group $G$ we can define the lower-central series as follows:\[G:=\gamma_0(G)\unrhd\gamma_1(G)\unrhd\cdots\unrhd\gamma_k(G)\cdots\] Where $\gamma_i(G)=\left[\gamma_{i-1}(G), G\right], i > 1$. Let $x,y$ be elements of a group $G$, we follow the usual definition of a commutator as $[x , y ] =x^{-1}y^{-1}xy$ . For two subgroups, $H$ and $K$ of $G$ we define $[H, K ] = \langle [h, k]\;|\;h\in H, k\in K \rangle$. If
the central series stops at identity, then we call that group a nilpotent group. If the length of the series is $c$, i.e., $\gamma_{c+1}(G)=1$, then we call it a nilpotent group of class $c$. This $c$ is also often referred to as the nilpotency class of a group or simply the class. In this paper we refer the nilpotent class simply as the class. It is not hard to show that if $G$ is a group of class $c$, then any commutator $[x_1 , x_2, ... , x_{c+1} ]$ of weight $c + 1$ is the identity.

Nilpotent groups of class 2 have many properties similar to that of abelian groups. We state without proof a well known lemma about groups of class $2$.
\begin{lemma}
Let $x,y\in G$ and assume that both $x$ and $y$ commute with $[x,y]$ then:
\begin{itemize}
\item[\emph{a)}] $[x,y]^n=[x^n,y]=[x,y^n]$ for all integer $n$
\item[\emph{b)}] $(xy)^n=x^ny^n[y, x]^{\frac{n(n-1)}{2}}$ for all $n\geq 0$
\end{itemize}
\end{lemma}
For a proof see Rotman~\cite[Lemma 5.42]{rotman}. The above lemma is a restatement of the fact, that in a nilpotent group of class 2, the map $x\mapsto[a,x]$ for a fixed $a$ is a linear map. This gives rise to the fact that $(x , y )\mapsto[x,y ]$ is a bilinear map from $G\times G\rightarrow G$. This is the central idea that we are going to use next. However, at this point we are obliged to report that in the case of groups of class 2, the bilinear map is from the group to the same group. Unlike the case of elliptic curves, where the bilinear map is from the group of an elliptic curve to a finite field. This change can have a profound effect on bilinear cryptography, especially in designing protocols.

\section{The central idea}
Let $G$ be a group of nilpotency class $2$. As discussed earlier, there are three users -- $\mathcal{A}$, $\mathcal{B}$ and $\mathcal{C}$ with private exponent $\alpha$, $\beta$ and $\gamma$ respectively. The main formula on which this key-exchange protocol is based an identity in a nilpotent group of class 2. 
\begin{equation}
 [x , y ]^n = [x^n , y ] = [x , y^n]\;\;
\end{equation}

As with the tripartite Diffie-Hellman key exchange, the users $\mathcal{A}$ and $\mathcal{B}$ and
$\mathcal{C}$ transmits in public $x^\alpha$, $y^\alpha$ ; $x^\beta$, $y^\beta$ and $x^\gamma$ , $y^\gamma$ respectively in public. 

\textbf{Key Exchange}. The tripartite key-exchange is as follows:
On receiving $x^\beta$ , $y^\gamma$ through the public channel, the user $\mathcal{A}$ can compute
$[x^\beta , y^\gamma ]^\alpha=[x , y ]^{\alpha\beta\gamma}$.
On receiving $x^\alpha$, $y^\gamma$ through the public channel, the user $\mathcal{B}$ can compute
$[x^\alpha , y^\gamma ]^\beta= [x , y ]^{\alpha\beta\gamma}$.
On receiving $x^\alpha$, $y^\beta$ through the public channel, the user $\mathcal{C}$ can compute
$[x^\alpha , y^\beta ]^\gamma = [x , y ]^{\alpha\beta\gamma}$.
The common key is $[x , y ]^{\alpha\beta\gamma}$.

\subsection{The primary security concerns} The primary security concerns are as follows:
\begin{itemize}
\item Given $x$ and $x^\alpha$ find $\alpha$. The same can be said for $y$. This is the
classic discrete logarithm problem.
\item From the information $x^\alpha$ and $y$, one can compute $[x^\alpha , y ] = [x , y ]^\alpha$.
Then it turns out to be the discrete logarithm problem in $[x , y ]$.
\item Note that $[x^\alpha, y^\beta ] = [x , y ]^{\alpha\beta}$. Clearly $[x , y ]^{\alpha\beta}$can be easily computed and so is $[x , y ]^\gamma$ from $[x , y^\gamma]$. The key-exchange is also broken, if we can compute $[x , y ]^{\alpha\beta\gamma}$ from $[x , y ]^{\alpha\beta}$ and $[x , y ]^\gamma$. This
is the classic Diffie-Hellman problem, also known as the computational Diffie-Hellman problem.
\end{itemize}

Bilinear cryptography described in previous section works in any nilpotent group of class 2. It is well known that every finite $p$-group is nilpotent. There are plenty of finite $p$-groups. So, it is a natural choice to investigate $p$-groups of class 2 for the purpose of bilinear cryptography. From now on we shall be  concerned with finite $p$-groups of class 2. In order to build an effective and secure bilinear cryptosystem using finite $p$-groups of class 2, users needs to choose the private exponents cleverly. Here we shed some light on the choices of private exponents.  
\paragraph{Exponent Semigroup}
Let $G$ be a group, define $\mathcal{E}(G)=\{n \in \mathbb{Z} \;|\; (xy)^n=x^ny^n \; \text{for all} \;x, y \in G\}$.  The semigroup $\mathcal{E}(G)$ is called the exponent semigroup and is of independent interest in group theory, see ~\cite{exp1,moravec}.
We state without proof a proposition which describes the structure of $\mathcal{E}(G)$. 
\begin{proposition} \emph{\cite[Proposition 3.2]{moravec}} Let $G$ be a finite $p$-group, $|G|=p^m$ and \emph{exp($G/Z(G)$)=$p^e$}. Then there exist a nonnegative integer $r$ such that $\mathcal{E}(G)=p^{e+r}\mathbb{Z}\cup(p^{e+r}\mathbb{Z}+1)$.
\end{proposition}
Let $G$ be a finite $p$-group of class 2 and $|G|=p^m$. Let $x, y \in G$ such that ord($x$)=$p^i$, ord($y$)=$p^j$ where $1< i, j \leq m$. The private exponents $\alpha, \beta, \gamma$ used in tripartite key-exchange protocol are independent of each other. In order to have a successful key-agreement it is necessary to have  $[x, y]^{\alpha\beta\gamma}\neq1$. Thus, $\alpha, \beta, \gamma$ must be choosen relatively prime to $p$, otherwise we could have $[x, y]^{\alpha\beta\gamma}=1$. Henceforth, we can assume that the private exponents $\alpha, \beta, \gamma < p^m$ and are relatively prime to $p$. Recall that $(xy)^n=x^ny^n[y, x]^{\frac{n(n-1)}{2}}$ for all $n\geq 0$. If $[y, x]^{\frac{n(n-1)}{2}}=1$ for all $x, y \in G$, implies $n\in \mathcal{E}(G)$. Thus, by above proposition $n=ap^{e+r}$ or $n=ap^{e+r}+1$, where $1 \leq a< p^{m-e-r}$. 
Therefore, we must avoid choosing private exponents $\alpha,\beta$ or $\gamma$ in $\mathcal{E}(G)=p^{e+r}\mathbb{Z}\cup(p^{e+r}\mathbb{Z}+1)$. If all $\alpha,\beta,\gamma$ is in $\mathcal{E}(G)$, then $\alpha\beta\gamma$ belongs to $\mathcal{E}(G)$ and by computing integers of the form $ap^{e+r}$ or $ap^{e+r}+1$ where $1 \leq a< p^{m-e-r}$ recover the private key $\alpha\beta\gamma$. However, this attack is not of much concern because $\alpha,\beta$ and $\gamma$ are chosen independent of one another and is a secret, so it is not likely that all three of these will belong to $\mathcal{E}(G)$. When there is no prior information on whether $\alpha,\beta$ and $\gamma$ belongs to $\mathcal{E}(G)$ is available, there is no guarantee that $\alpha\beta\gamma$ will belong to $\mathcal{E}(G)$ and the attack then is just a mere exhaustive search.

We now hope that there is a convincing argument that our central idea runs parallel to the pairing based cryptosystems currently being studied. There is a lot that can be said about protocols using the above idea. A lot can be said about "provable" or semantic security of those cryptosystems. It can be an active field of study to design proper protocols using the above idea in an appropriate security model. However, this paper is not about "provable" or semantic security. It is about finding the right group in which the above mentioned scheme
works nicely and securely. As we know the security of the discrete logarithm problem depends on the presentation of the group. There are three most commonly used presentations of finite groups.
\begin{itemize}
\item Permutation presentation.
\item Polycyclic presentation.
\item Matrix presentation.
\end{itemize}

\section{Finding a right group}
From the above discussion it is clear that the stepping stone to bring this
idea to light is to look at 2-generator $p$-groups of nilpotency class 2. Fortunately there is a lot known about 2-generator $p$-groups of class 2, these
groups have even been classified. The automorphism group
of these groups are somewhat known but there is no mention in the literature
on the linear representations of these groups.
Our idea simply is to find suitable representations of these 2-generator
$p$-groups of class 2 and then use it in the ideas described above. In particular, for the purpose of an exposition,  we are interested in the 
extra-special $p$ groups.
\subsection{Extra-special $p$-groups}
Let $G$ be a finite $p$-group. Then $G$ is defined to be special if either $G$ is
elementary abelian or $G$ is of class 2 and $G' = \Phi(G) = Z (G)$ is elementary abelian. If $G$ is
a non-abelian special group with $|Z(G)| = p$, then $G$ is said to be extraspecial. For example,
dihedral group $D_{8}$ and quaternion group $Q_{8}$ are extraspecial. An example of our interest is the following: 
For $\alpha \geq \beta \geq \gamma \geq 1$, 
\[
 G=\langle a,b|a^{p^{\alpha}}=b^{p^{\beta}}=[a,b]^{p^{\gamma}}=1,[a,b,a]=[a,b,b]=1\rangle.
\]
It is clear that $G$ is a two-generator $p$-group of nilpotency class 2. It is not hard to see that the derived subgroup $G^\prime$ is 
cyclic and of order $p^{\gamma}$. Furthermore $G/G^\prime$ is isomorphic to $C_{p^{\gamma}}\oplus C_{p^{\beta}}.$ Here $C_{n}$ is the 
cyclic group of order $n$.

Here our main goal is to find suitable linear representations for the above mentioned group. The following theorem
will be very useful for our study. 
\begin{theorem}
 Every extraspecial $p$-group $G$ is the central product of extraspecial groups of order $p^3$. Irreducible representations of $G$
 are all obtained as tensor product of irreducible representations of the individual factors of $G$.
\end{theorem}
We now give some examples of $p$-groups that we will use for our study.
The modular $p$-group $Mod_{n}(p)$ is given by the generators and relations
\begin{equation}
 Mod_{n}(p)=\langle a,b | a^{p^{n-1}}=b^p=1, a^b=a^{1+p^{n-2}}\rangle.
\end{equation}
And the group 
\begin{equation}
 M(p)=\langle a,b | a^p=b^p=[a,b]^p=1,[a,b,a]=[a,b,b]=1\rangle.
\end{equation}
In light of the above theorem we will study the linear representations of extraspecial $p$-groups of order $p^3$. The following 
theorem characterize the non-abelian groups of order $p^3$.
\begin{theorem}
 A non-abelian $p$-group $G$ of order $p^3$ is extraspecial and is isomorphic to one of the groups $Mod_{3}(p)$, $M(p)$, $D_{8}$
 or $Q_{8}$. 
\end{theorem}
For the purpose of bilinear cryptography we will look for the irreducible faithful linear representations of $Mod_{3}(p)$
over a finite field. As we know that~\cite[Theorem 5.5]{goren}, if $G$ be an extraspecial $p$-group of order $p^{2r+1}$ and  $F$ be a field of 
characteristic 0 or prime to $p$ which contains a primitive $p^2$-root of unity. Then the faithful representation of $G$ over $F$
are all of degree $p^r$.
Throughout we will assume that $p$ and $s$ are odd primes and $q=s^{r}$. Let $\zeta$ be a primitive $m^{th}$ root of unity in
${\overline{\mathbf{F}}_{s}}$ and let $k$ be a positive integer with $|k|_{m}=n$. Where $|k|_m$ is the order of $k$ in the multiplicative groups of 
units mod $m$.
\begin{lemma} \emph{\cite[Lemma 2.2]{che}}
The Frobenius automorphism $\tau$ on ${\overline{\mathbf{F}}_{s}}$ defined by $\tau (x)=x^{q}$ permutes the elements of
$\{\zeta, \zeta^{k},\cdots,\zeta^{q^{n-1}}\}$ iff $q\equiv k^{i}\pmod{m}$ for some $0\leq i\leq n-1$.
\end{lemma}
\begin{theorem} \emph{\cite[Theorem 2.5]{che}}
 Let $G=\langle a,b | a^m=1=b^n, b^{-1}ab=a^k\rangle$ where $|k|_{m}=n$. Let $\rho$ be the representation of $\langle a \rangle$ 
 defined by $\rho (a)=\zeta$ where $\zeta$ is a primitive $m^{th}$ root of unity in  $\overline{\mathbf{F}}_{s}$, $(s,m)=1$. Then the
 induced representation $\rho^{G}$ is realizable over  $\mathbf{F}_{q}$ iff $q\equiv k^{i}\pmod{m}$  for some $0\leq i\leq n-1$.
\end{theorem}
 Observe that if we choose $m=p^2$, $n=p$, and $k=p+1$ in the above theorem then $G=Mod_{3}(p)$ as $(p+1)^p\equiv 1\pmod{p^2}$.
 Now as we know that by~\cite[Proposition 2.1]{che} $\rho ^{G}$ is irreducible over ${\overline{\mathbf{F}}_{s}}$ iff $|k|_{m}$ is equal to the order of $b$.
 Observe that $\zeta, \zeta^{q},\cdots, \zeta^{q^{n-1}}$ are distinct and $\zeta^{q^{n}}=\zeta$ whence $|q|_{p^2}=p$. 
 By ~\cite[Lemma 2.2]{che}, $q\equiv (p+1)^{i}\pmod{p^2}$ for some $i$. As $|q|_{p^2}=|p+1|_{p^2}$ implies that $i$ and $p$ are co-prime.
 Hence, letting $c=b^i$ we have $G=\langle a,c | a^m=1=c^n, c^{-1}ac=a^q\rangle$. The induced representation $\rho ^{G}$ using coset 
 representatives $1,c,c^2, \cdots, c^{p-1}$ is given by 
 \begin{eqnarray*}
\rho^{G}(c)=
 \begin{pmatrix}
  0 & 0&\cdots&0&1 \\
  1 &0&\cdots&0&0 \\
  0&1&\cdots &0&0\\
  \vdots &\vdots &\cdots &\vdots &\vdots\\
  0&0&\cdots &1 &0
 \end{pmatrix}
 \end{eqnarray*}
 \begin{eqnarray*}
 \quad \rho^{G}(a)=
 \begin{pmatrix}
 \zeta & &&&0 \\
 &\zeta^{q}& \\
  & && \ddots&\\
  0&&&&\zeta^{q^{p-1}}
 \end{pmatrix} 
\end{eqnarray*}
Again by~\cite[Lemma 2.2]{che} we can see that $V^{-1}\rho^{G}(a)V=X \quad \textrm{and} \quad V^{-1}\rho^{G}(c)V=Y \in M_{p}(\mathbf{F_{q}})$ 
where $V$ and $X$(the companion matrix of $f(x)=\displaystyle \prod_{i=1}^{p-1}(x-\zeta^{q^i})$)
are as follows: 
\begin{eqnarray*}
V=
 \begin{pmatrix}
  1 &\zeta_{0} &\cdots&\zeta_{0}^{p-1} \\
  1 &\zeta_{1} &\cdots&\zeta_{1}^{p-1} \\
  \vdots&\vdots&\cdots &\vdots\\
  1&\zeta_{p-1}&\cdots &\zeta_{p-1}^{p-1}
 \end{pmatrix}, 
 \end{eqnarray*}
 \begin{eqnarray*}
 \quad X=
 \begin{pmatrix}
 0&0&\cdots &0&-a_{0}\\
 1&0&\cdots &0&-a_{1} \\
 0 &1&\cdots&0&-a_{2}\\
 \vdots&\vdots&\cdots &\vdots&\vdots\\
  0&0&\cdots&1&\zeta^{q^{p-1}}
 \end{pmatrix}, 
 \end{eqnarray*}
 \begin{eqnarray*}
 \quad Y=
 \begin{pmatrix}
 1 & &&& \\
 &\zeta^{p}& \\
  & && \ddots&\\
  &&&&\zeta^{(p-1)p}
 \end{pmatrix} 
\end{eqnarray*}
where $\zeta_{i}=\zeta^{q^i}$, $0\leq i\leq n-1$.
Here we have $f(x)=x^p-\zeta^p$, hence we get the equivalent representation $\sigma$ of $\rho^G$ which is given by: 

\begin{eqnarray*}
\sigma(a)=
 \begin{pmatrix}
  0 & 0&\cdots&0&\zeta^p \\
  1 &0&\cdots&0&0 \\
  0&1&\cdots &0&0\\
  \vdots&\vdots&\cdots &\vdots&\vdots\\
  0&0&\cdots &1 &0
 \end{pmatrix}, 
 \end{eqnarray*}
 \begin{eqnarray*}
 \quad \sigma(c)=
 \begin{pmatrix}
 1 & &&& \\
 &\zeta^{p}& \\
 &&\zeta^{2p}&\\
  & && \ddots&\\
  &&&&\zeta^{(p-1)p}
 \end{pmatrix} 
\end{eqnarray*} 
Note that  $\sigma(ac)=
  \begin{pmatrix}
  0 & 0&\cdots&0&1 \\
  1 &0&\cdots&0&0 \\
  0&\zeta^p&\cdots &0&0\\
  \vdots&\vdots&\cdots &\vdots&\vdots\\
  0&0&\cdots &\zeta^{(p-2)p} &0
 \end{pmatrix}$\\
The characteristic polynomial of $\sigma(a)$ and $\sigma(ac)$ is same which is $x^p-\zeta^p$.\\
Now consider the tensor product of representation $\sigma$ with itself. If the characteristic polynomials of 
$A$ and $B$ factor as $P_{A}(x)=\displaystyle \prod_{i=1}^{n}(x-\lambda_{i})$ and $P_{B}(x)=\displaystyle \prod_{i=1}^{m}(x-\mu_{j})$,
then the characteristic polynomial of $A\otimes B$ is $P_{A\otimes B}(x)=\displaystyle \prod_{i=1}^{n}\displaystyle \prod_{i=1}^{m}
(x-\lambda_{i}\mu_{j})$. Thus characteristic polynomial of $\sigma(a)$, $\sigma(ac)$ is $(x^p-\zeta^p)^p$. Observe that the polynomial
$(x^p-\zeta^p)$ is irreducible over $F$ where $F$ is a subfield of $F_{q}$ with $\zeta^{p}\in F$.

\section{Conclusion}
In this short note, we introduced a novel idea of pairing based cryptosystem using matrices. The idea is simple, use linear representation of groups of class 2. As a test case we looked at the extraspecial $p$-group of exponent $p^2$. For this particular group, the set of parameters is not encouraging. From~\cite{che}, it follows that the only possible matrices that we can come up with is matrices of size $p$ over $\mathbb{Z}_p$. The most security this can provide is the security in the finite field $\mathbb{F}_{p^p}$. However, since the size of the field and the size of the matrix must be the same, this is of little practical value. The size of the matrix is too large. However, we hope, that there are other groups of class 2 for which we can get better parameters.
\bibliography{AAECC}
\bibliographystyle{splncs03}
\end{document}